\documentclass[aip,jcp,reprint,superscriptaddress]{revtex4-1}
\pdfoutput=1
\usepackage{graphicx}
\usepackage{amssymb}
\usepackage{amsmath}
\newcommand{\comment}[1]{}
\usepackage{textcomp}
\usepackage{hyperref}
\usepackage{xcolor}
\usepackage{adjustbox}

\newcommand{\sub}[1]{\ensuremath{_{\mathrm{#1}}}}
\renewcommand{\vec}[1]{\mathbf{#1}}

\begin{document}
 
\title{First-principles electrostatic potentials for\\reliable alignment at interfaces and defects}

\author{Ravishankar Sundararaman}
\email{sundar@rpi.edu}
\affiliation{These authors contributed equally.}
\affiliation{Department of Materials Science and Engineering, Rensselaer Polytechnic Institute, Troy, NY 12180}
\author{Yuan Ping}
\email{yuanping@ucsc.edu}
\affiliation{These authors contributed equally.}
\affiliation{Department of Chemistry and Biochemistry, University of California, Santa Cruz, CA 95064}

\begin{abstract}
Alignment of electrostatic potential between different atomic configurations is necessary for 
first-principles calculations of band offsets across interfaces and formation energies of charged defects.
However, strong oscillations of this potential at the atomic scale make alignment challenging,
especially when atomic geometries change considerably from bulk to the vicinity of defects and interfaces.
We introduce a method to suppress these strong oscillations
by eliminating the deep wells in the potential at each atom.
We demonstrate that this method considerably improves the system-size convergence
of a wide range of first-principles predictions that depend on alignment of
electrostatic potentials, including band offsets at solid-liquid interfaces,
and formation energies of charged vacancies in solids and at solid surfaces in vacuum.
Finally, we use this method in conjunction with continuum solvation theories
to investigate energetics of charged vacancies at solid-liquid interfaces.
We find that for the example of an NaCl (001) surface in water,
solvation reduces the formation energy of charged vacancies by 0.5 eV:
calculation of this important effect was previously impractical due to
computational cost in molecular-dynamics methods.
\end{abstract}

\maketitle

Electrostatic potential alignment plays a central role in determining interfacial band offsets
and charged defect energetics\cite{NMinterface,FreysoldtRMP,Ping2015,BandOffsets}
in first-principles calculations based on Kohn-Sham density functional theory (DFT).
Specifically, in infinite systems treated with periodic boundary conditions,
electrostatic potentials, and consequently the energy eigenvalues, are defined only up to
an undetermined constant. Typically the average potential in the unit cell is set to zero.
In interfacial band-offset calculations, this necessitates aligning electrostatic potentials
of bulk calculations with those of the bulk-like regions in the interface calculation.\cite{Ping2015,BandOffsets}
For charged systems, total energies are also sensitive to the arbitrary absolute offset of the electrostatic potential.
In evaluating formation energies of charged defects, this is addressed by aligning electrostatic potentials
far away from the defect with that of the pristine system (without the defect).\cite{stephan2009,walle20014}

A critical issue that affects all these calculations is that electrostatic potentials in DFT
oscillate strongly with magnitudes of tens of eV, due to deep wells at the location of each atom.
This makes alignment of potentials difficult and necessitates large computationally-expensive supercells.
In this paper, we address this issue by a simple redefinition of electrostatic potentials in DFT
that eliminates the deep wells centered at each atom and brings the potentials to the eV scale,
as detailed in Section~\ref{sec:Potential}.
We then show that this change considerably improves the supercell-size convergence
of band offsets across solid-liquid interfaces in Section~\ref{sec:BandOffsets},
and of formation energies of charged defects in solids in Section~\ref{sec:BulkDefects}
and at solid surfaces in Section~\ref{sec:SurfaceDefects}.
Our redefinition of the electrostatic potential thereby enhances the computational
efficiency of first-principles predictions for a wide range of properties.
This makes previously unexplored properties now accessible, such as the energetics of
charged defects at solid-liquid interfaces demonstrated in Section~\ref{sec:SurfaceDefects}.

\section{Electrostatic potential}\label{sec:Potential}

In plane-wave pseudopotential DFT calculations, the total electrostatic potential,
including contributions due to both electrons and nuclei, is typically defined as
\begin{equation}
V(\vec{r})
= \underbrace{\int \mathrm{d}\vec{r}' \frac{n(\vec{r}')}{|\vec{r}-\vec{r}'|}}_{V_H(\vec{r})}
+ \underbrace{\sum_i V\sub{loc}^{(i)}(|\vec{r}-\vec{r}_i|)}_{V\sub{loc}(\vec{r})}.
\label{eq:Vorig}
\end{equation}
Above, $V_H(\vec{r})$ is the Hartree potential due to the valence electron density $n(\vec{r})$,
and $V\sub{loc}(\vec{r})$ is the local part of the pseudopotential
that includes the potential due to the nuclei and the core electrons,
written as sum of spherical functions $V\sub{loc}^{(i)}$
centered at each nucleus located at $\vec{r}_i$.

\begin{figure}
\includegraphics[width=\columnwidth]{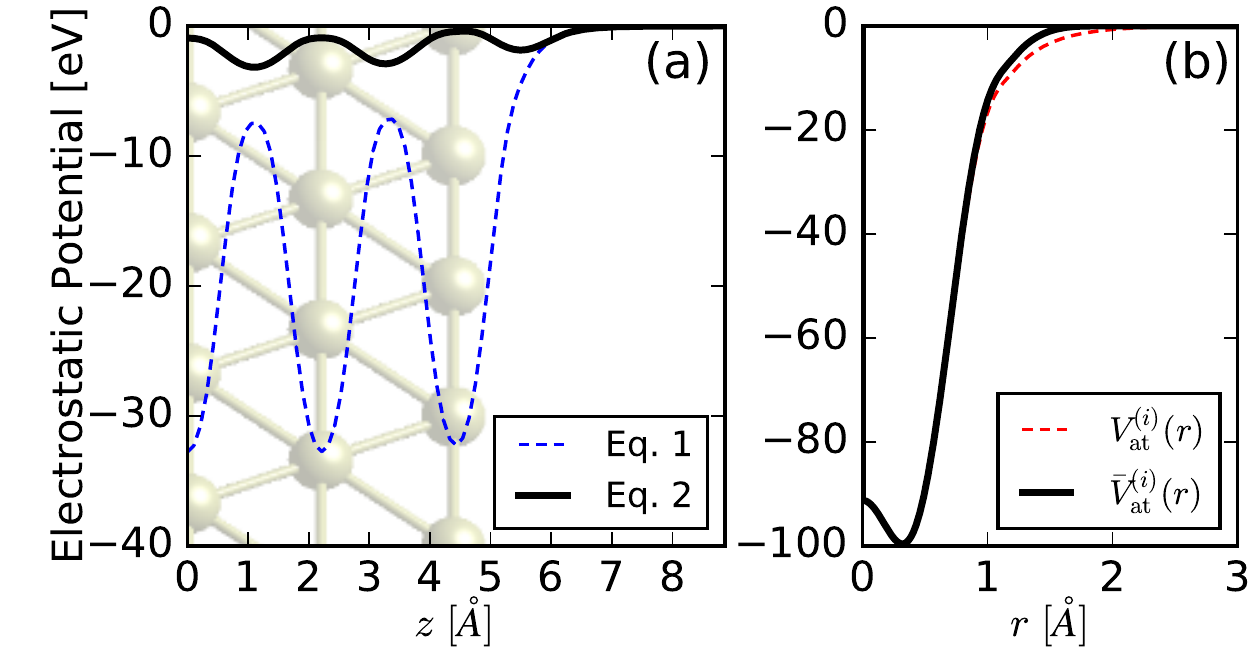}
\caption{
(a) Comparison of the conventional (Eq.~\ref{eq:Vorig}) and revised (Eq.~\ref{eq:Vnew})
planarly-averaged electrostatic potential for a 5-layer Ir(111) slab calculation.
Note the drastic reduction of oscillations from the Hartree to eV scale.
(b) Comparison of the original ($V\sub{at}^{(i)}(r)$) and pseudized
($\bar{V}\sub{at}^{(i)}(r)$) electrostatic potential of an Ir atom.
\label{fig:Vatom}}
\end{figure}

This potential exhibits deep wells at each nucleus
making averages of the potential highly oscillatory.
For example, Fig.~\ref{fig:Vatom}(a) shows the planarly-averaged
electrostatic potential from a DFT calculation of an Ir(111) surface,
which shows the characteristic deep wells at each (111) plane.
Differences in the electrostatic potential between two calculations would cancel these
deep wells provided the atoms in the two calculations can be aligned exactly,
which is usually not possible when the ionic geometry is optimized.
Consequently, all applications of DFT involving comparisons of
electrostatic potential between systems with optimized ionic geometries
require large enough unit cells that at least some of the atoms move
negligibly and do not induce appreciable oscillations in the potential.
This paper presents a method to minimize the magnitude of such oscillations,
allowing more efficient and accurate smaller-unit-cell calculations for
a number of such applications, including interfacial band alignment
and charged defect formation energies.

Note that $V(\vec{r})$ is not directly a physically meaningful property.
It depends implicitly on the pseudopotential through the separation of
electrons between core and valence, and through the formulation of $V\sub{loc}(r)$.
The absolute values of $V(\vec{r})$, as well as its averages in unit cells,
depend on these choices.  Only the difference of potential between two regions with a
similar atomic composition (calculated using the same pseudopotentials) is physically meaningful
eg. between bulk-like regions of an interface calculation with corresponding bulk calculations.
All applications of electrostatic potentials, such as band offset or charged-defect
formation energy calculations, always involve such differences in the final physical predictions.
Consequently, any modification of the potential within the core region of atoms,
if done consistently for each atom type in all involved calculations,
will not change the converged values of physical properties.
We exploit this degree of freedom in the definition of $V(\vec{r})$ to minimize
its oscillations and improve cell-size convergence in its applications.

Specifically, we eliminate the deep well in the potential at each nuclear position
by subtracting the potentials of neutral atoms $\bar{V}\sub{at}^{(i)}$
centered at each nucleus, defining the \emph{revised} electrostatic potential
\begin{equation}
\bar{V}(\vec{r}) = V_H(\vec{r}) + V\sub{loc}(\vec{r}) - \sum_i \bar{V}\sub{at}^{(i)}(|\vec{r}-\vec{r}_i|).
\label{eq:Vnew}
\end{equation}
Effectively, this always defines the electrostatic potential of each calculation
as a difference between two systems evaluated with atoms located at the same positions.
(Formally, the second system is a collection of neutral atoms
at the same positions that do not interact with each other.)

The electrostatic potential of neutral atoms $V\sub{at}^{(i)}$ is spherically symmetrical,
exponentially decays away from the atom and can be easily and rapidly evaluated on a radial grid.
It consists of the same two terms as Eq.~\ref{eq:Vorig}: $V\sub{loc}^{(i)}(r)$ which is directly
stored in atom pseudopotentials, and the Hartree potential which can be computed from the
electron density by radial integration of the Poisson equation.

However, $V\sub{at}^{(i)}$ is not always localized to individual atoms,
since the exponential decay of the atomic electron density can be slow.
Consequently, subtracting $V\sub{at}^{(i)}$ would have the undesirable
side effect of modifying the potential outside the atoms,
where the magnitude of the potential is already
small and such a subtraction is not necessary.
We therefore instead subtract a pseudized version, $\bar{V}\sub{at}^{(i)}$,
of the neutral atom potential $V\sub{at}^{(i)}$, which is constructed
to match $V\sub{at}^{(i)}$ in the core regions of the atom
but smoothly approaches zero outside a cutoff radius $R_i$.
Specifically we require that the revised potential preserves
the value, first and second derivatives at $r=0$, and vanishes
at $R_i$ smoothly with zero first and second derivatives.
The form of the revised potential with the lowest order
polynomial modification that achieves this is
\begin{equation}
\bar{V}\sub{at}^{(i)}(r) = 
\begin{cases}
V\sub{at}^{(i)}(r) - \frac{r^3}{R_i^3} \left( a_3 + a_4\frac{r}{R_i} + a_5\frac{r^2}{R_i^2} \right), & r<R_i \\
0, & r \ge R_i,
\end{cases}
\end{equation}
with the coefficients given by
\begin{equation}
\left(\begin{array}{c}
a_3\\
a_4\\
a_5
\end{array}\right)
=
\left(\begin{array}{ccc}
 10 & -4 & 1/2 \\
-15 &  7 & -1  \\
  6 & -3 & 1/2
\end{array}\right)
\cdot
\left(\begin{array}{c}
V\sub{at}^{(i)}(R_i)\\
R_i\frac{dV\sub{at}^{(i)}}{dr}\big|(R_i)\\
R_i^2\frac{d^2V\sub{at}^{(i)}}{dr^2}\big|(R_i)\\
\end{array}\right)
\end{equation}
to satisfy the smoothness constraints at $r=R_i$.

The cutoff radius $R_i$ is arbitrary, as long as it is small enough to
avoid substantial overlap between neighbouring atoms, and it is large enough
to encompass the core region so that it can eliminate the deep well in the potential.
For definiteness, we follow the prescription for the vdW radius used in the DFT-D2
dispersion correction functional,\cite{Dispersion-Grimme} and set $R_i = 1.1 \times$
the radius at which the electron density crosses $10^{-2} a_0^{-3}$.
This also has the advantage of determining the radius from the electron density,
and not requiring a tabulation of atomic radii for all elements.

Fig.~\ref{fig:Vatom}(b) shows that the pseudized potential for an Iridium atom
(using the GBRV ultrasoft psuedopotential\cite{GBRV} for Ir), mostly follows
the atom potential with only a small modification that forces it to zero at finite radius.
Using this pseudized atom potential in Eq.~\ref{eq:Vnew}
then reduces the oscillations in the electrostatic potential
by an order of magnitude relative to the original Eq.~\ref{eq:Vorig},
as shown for the Ir(111) surface example in Fig.~\ref{fig:Vatom}(a).
The remainder of this paper shows that this reduction in the magnitude
of the electrostatic potential oscillations considerably simplifies
a number of DFT applications involving the potential, including
band alignment and charged defect calculations.

\section{Computational details}

We implement the above method and perform all calculations here
using the open-source plane-wave DFT software JDFTx.\cite{JDFTx}
We use the Perdew-Burke-Ernzerhoff generalized-gradient approximation\cite{PBE}
to the exchange-correlation functional, and ultrasoft pseudopotentials
from the `GBRV' set\cite{GBRV} at the recommended kinetic energy cutoffs
of 20~$E_h$ for the wavefunctions and 100~$E_h$ for the charge density.
We use Monkhorst-Pack\cite{monkhorst1976} $k$-point grids for Brillouin zone sampling,
with the number of $k$-points per dimension chosen to have a minimum
supercell length of 20~\AA\ in each direction.
For all surface calculations (slab geometry), we use truncated
Coulomb potentials to eliminate interactions between periodic images
of the slab across the vacuum (or liquid) region.\cite{TruncatedEXX}

\section{Band offsets at solid-liquid interfaces}\label{sec:BandOffsets}
The electrostatic potential profile plays a central role in
determining the energy level alignment across interfaces,
which in turn affects the charge transport across interfaces.
DFT calculations of the electrostatic potential
are routinely used to calculate the energy level alignment across
solid-solid interfaces, such as the Schottky barrier height
in metal-semiconductor interfaces.\cite{Baldereschi1988,Baldereschi1998}
Such calculations require alignment of electrostatic potentials between
bulk materials and the corresponding bulk regions in an explicit model
of the interface in order to account for the interfacial dipole effects.
Even in theories beyond DFT, such as many-body perturbation using the GW method
that improves accuracy for electronic states and band energies,\cite{2015marco}
these electrostatic potential shifts are typically evaluated using
DFT.\cite{Anh2014,Ping2015,pingwo32014,review2013,Hinuma2014}

Similar calculations for solid-liquid interfaces are considerably
more challenging, requiring \emph{ab initio} molecular dynamics
to sample several thousands of configurations of the liquid.\cite{Anh2014,NMinterface}
Solvation models, which directly replace the thermodynamically-averaged
effect of the liquid with that of a continuum dielectric,
can substantially simplify such calculations for solid-liquid interfaces
by eliminating the need for sampling liquid configurations.
Recent solvation models\cite{SaLSA,CANDLE} can accurately predict the band alignment
at these interfaces in comparison to experimental measurements or
\emph{ab initio} molecular dynamics simulations, at a fraction of the effort.\cite{BandOffsets}

\begin{figure}
\includegraphics[width=\columnwidth]{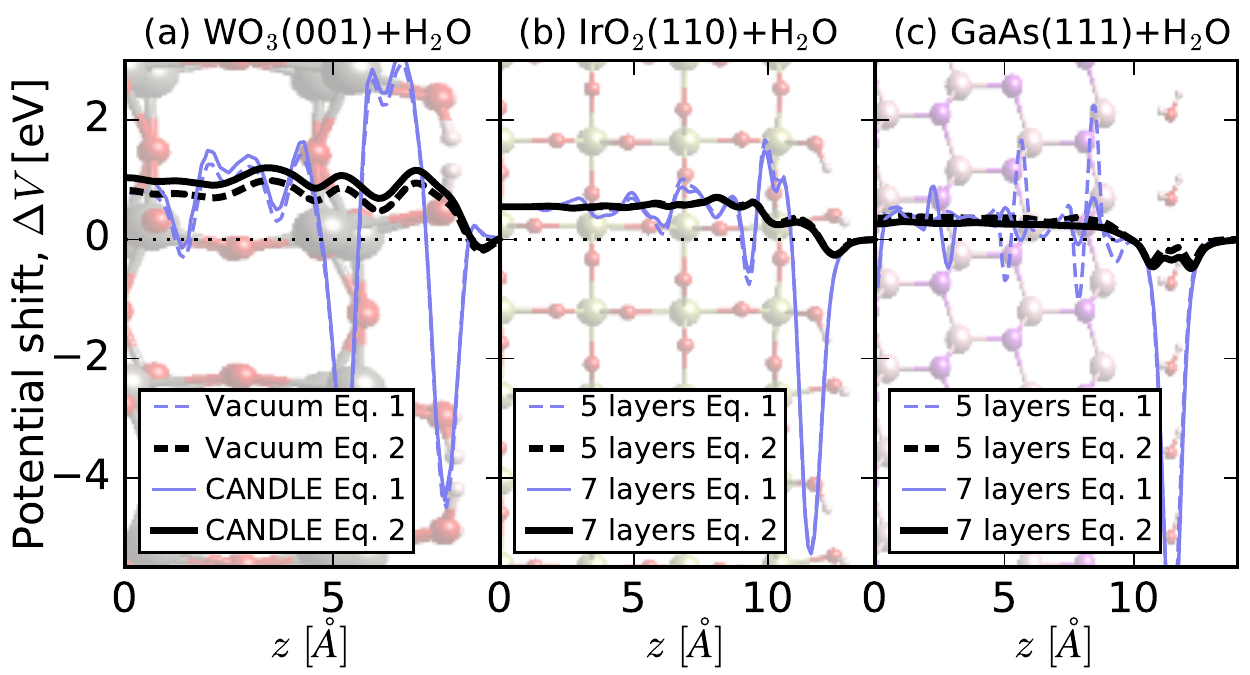}
\caption{Comparison between electrostatic potential shifts ($\Delta V$)
predicted using original (Eq.~\ref{eq:Vorig}) and revised (Eq.~\ref{eq:Vnew}) schemes for
the addition of an explicit monolayer of water molecules to
(a) $\gamma$-monoclinic WO$_3$ (001) surface in vacuum and
with the CANDLE solvation model,\cite{CANDLE} and to (b) IrO$_2$ (110)
and (c) GaAs (111) surfaces described using 5 and 7 layer slabs.
Surface structures shown in the background are to scale with the $x$-axis.
Substantial oscillations in $\Delta V$ in the original scheme
are suppressed by our revised scheme (Eq.~\ref{eq:Vnew}), making the
potential converge to the bulk value in fewer layers.
\label{fig:BandOffset}}
\end{figure}

Our method for reducing electrostatic potential oscillations
due to nuclei is applicable to all these cases of band alignment 
at solid/solid, solid/liquid interfaces; here we will first
demonstrate the application for band edge shifts by using
one layer of explicit water molecules with and without solvation models.
Fig.~\ref{fig:BandOffset}(a) shows the shift of the electrostatic potential, $\Delta V$,
of a $\gamma$-monoclinic WO$_3$ (001) surface, modeled with an inversion-symmetric
4 layer slab, due to a layer of explicit water molecules alone (dashed lines)
and additionally with an implicit solvent model (solid lines).
(For the purposes of this comparison, we consider the stoichiometric surface
for simplicity; see Ref.~\citenum{BandOffsets} for a detailed analysis on
the important role of surface oxygen vacancies at this surface.)
These cases are shown when the potential is evaluated both
using the original Eq.~\ref{eq:Vorig} scheme (thinner blue lines)
and our revised Eq.~\ref{eq:Vnew} scheme (thicker black lines).
The strong binding of water molecules to the hydrophilic WO$_3$ surface
perturbs the surface structure substantially, and this leads to
strong oscillations in $\Delta V$ of magnitude over 2~eV extending into the
inner layers in the original scheme. With exactly the same perturbed geometries,
our revised scheme for calculating electrostatic potentials
reduces the overall magnitude of the oscillations to
within 0.1~eV, making the identification of the net band offsets
($\approx 0.8$~eV without and 1.0~eV with the solvation model) far clearer.

Similarly, Fig.~\ref{fig:BandOffset}(b) compares the $\Delta V$ induced by
one layer of explicit water molecules added to the IrO$_2$ (110) surface,
modeled with inversion-symmetric 5 and 7 layer slabs.
As before, the revised scheme reduces oscillations
from the $\sim 1$~eV scale to within the $0.1$~eV scale,
resulting in an essentially flat potential profile
beyond the first unit cell of the surface.
Note however that despite the strong oscillations in the original scheme,
the potential shift at the center of the slab ($z=0$) is not strongly
affected by the revised scheme; this is because the central layer atoms
do not change their $z$ positions on account of the inversion symmetry of our slab.
Indeed, Fig.~\ref{fig:BandOffset}(c) shows similar comparisons for a
polar GaAs(111) surface which necessarily breaks inversion symmetry.
In this case, the oscillations in the original scheme persist all the way
through to the center of the slab, making identification of the net band offsets challenging,
whereas our revised scheme (Eq.~\ref{eq:Vnew}) yields a flat $\Delta V$ profile beyond
the first layer enabling unambiguous determination of the band offset ($\approx 0.3$~eV).

\section{Formation energies of charged defects}\label{sec:BulkDefects}

Eliminating oscillations in the electrostatic potential improves system-size convergence,
and hence reduces computational effort, more generally wherever alignment of this potential matters.
We next demonstrate the efficacy of our method for a prominent example of increasing recent interest:
the calculation of the formation energies and charge transition levels of charged defects such as vacancies.\cite{PRLbulk,Komsa2013,kumagai2014}
Briefly, the formation energy of a charged vacancy $V_A^q$ of an atom $A$ with net charge $q$, is given by
\begin{equation}
E\sub{form} = E_{\mathrm{bulk+}V_A^q} - E\sub{bulk} + \mu_A - q E\sub{Fermi} + E\sub{corr},
\label{eqn:Eform}
\end{equation}
where the first two terms compare the energies of a supercell of the material
with and without the vacancy, while the third and fourth terms account for the
difference in numbers of atom $A$ and electrons respectively
with their corresponding chemical potentials. 
The fundamental challenge in calculating $E\sub{form}$ is that $E_{\mathrm{bulk+}V_A^q}$
converges very slowly with supercell size ($\propto L^{-1}$) due to
the periodic interactions between the charge $q$.\cite{FreysoldtRMP}
The final term $E\sub{corr}$ estimates and compensates for these interactions by
using a Gaussian charge model of the vacancy and calculating the difference
between the self energy of that charge in isolation and in the finite supercell,
in the background of the bulk dielectric constant of the material.\cite{PRLbulk}
(The relevant dielectric constant is the low-frequency value $\epsilon_0$ if the
atoms are optimized, and the optical value $\epsilon_\infty$ if the atoms are fixed.)
Additionally, it includes an `alignment' contribution $-q\Delta V$ due to the difference
in electrostatic potential of the model charge far from the defect
compared to that calculated by DFT. This is necessary because of the
indeterminacy of absolute potential in periodic boundary conditions as
we discussed earlier. See Ref.~\citenum{PRLbulk} for further details.

\begin{figure}
\includegraphics[width=\columnwidth]{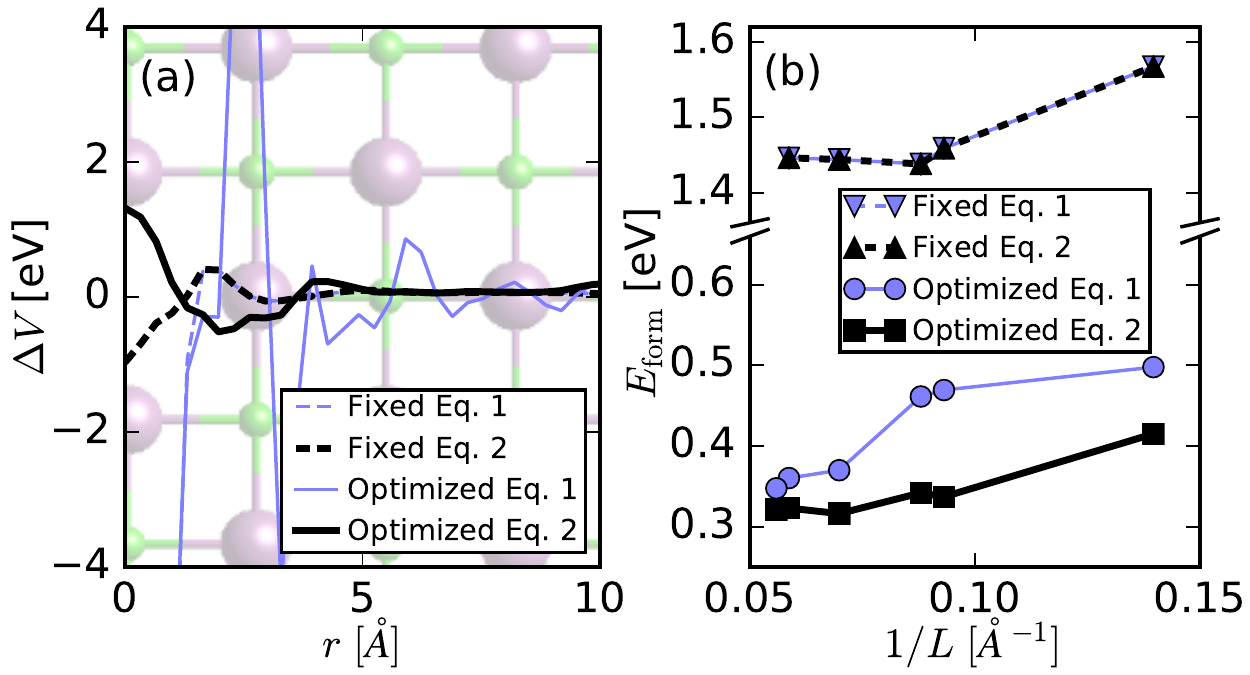}
\caption{(a) Alignment potential $\Delta V$ between DFT and Gaussian charge model
for a $q=+1$ Cl vacancy in NaCl, averaged radially away form the vacancy,
compared between Eq.~\ref{eq:Vorig} and \ref{eq:Vnew}, both for fixed and optimized atom positions.
The crystal structure is shown in the background to scale with respect to the $x$-axis.
(b) Corresponding convergence of the formation energy $E\sub{form}$ of the $q=+1$ Cl vacancy
with inverse supercell size, where $L = \Omega^{1/3}$ for supercell volume $\Omega$.
Reduced oscillations in $\Delta V$ using Eq.~\ref{eq:Vnew} enable accurate calculations of $E\sub{form}$
with much smaller supercells (larger $1/L$).
\label{fig:BulkVacancy}}
\end{figure}

Here, we apply this calculation method for the formation energy of $q=+1$ Cl vacancies in bulk NaCl,
and examine the effect of our revised scheme for calculating
electrostatic potentials on the alignment potential $\Delta V$.
Fig.~\ref{fig:BulkVacancy}(a) compares this potential, radially averaged from
the center of the defect, as evaluated using Eqs.~\ref{eq:Vorig} and \ref{eq:Vnew}, both when the
atom positions are fixed at their bulk values, and when they are optimized.
When the atoms are fixed (dashed lines), the original and revised
schemes agree exactly beyond 2~\AA, because the remaining atoms exactly overlay
and the subtraction scheme has no effect.
These results agree very well with calculations of Ref.~\citenum{Komsa2013} for the same system.

However, once the atoms are optimized, $\Delta V$ exhibits oscillations
with magnitude $\sim 0.5$~eV 5~\AA~away and $\sim 0.1$~eV 10\AA~away
even after radial averaging in the original scheme, while our revised scheme
yields a flat $\Delta V$ profile beyond $r = 5$~\AA.
Fig.~\ref{fig:BulkVacancy}(b) shows the corresponding predictions for
the formation energies using the aforementioned method.\cite{PRLbulk}
Note that optimizing the atom positions is extremely important here:
it lowers the vacancy formation energy by greater than 1~eV.
Geometry optimization was absent in previous work such as
Ref.\citenum{Komsa2013}, however, precisely due to the difficulty
in aligning electrostatic potentials with large oscillations
when the atom positions do not overlay exactly.
As expected from the $\Delta V$ comparison, the predicted formation energy
for optimized atom positions using Eq.~\ref{eq:Vorig} for the potential,
exhibits errors $\sim 0.1$~eV even for $1/L \sim 0.1\AA^{-1}$,
while the revised scheme gives more accurate results with
even smaller supercells of $1/L \sim 0.14\AA^{-1}$.
Furthermore, the results using Eq.~\ref{eq:Vnew} show a clear convergence for the
four larger supercells (smaller $1/L$ than $\sim 0.14\AA^{-1}$), while those of Eq.~\ref{eq:Vorig} do not.
Overall, Eq.~\ref{eq:Vnew} enables results with better accuracy using supercells
that are twice the $1/L$ i.e. that contain 1/8 as many atoms,
compared to what was previously possible with Eq.~\ref{eq:Vorig}.

\section{Charged defects at interfaces}\label{sec:SurfaceDefects}
Atoms at the surface of solid are usually less tightly bound than those in the bulk,
and surfaces are often more prone to contain defects such as vacancies, and these
can play an important role in determining interface potentials and band alignment.\cite{BandOffsets}
Calculating the formation energy of charged defects at surfaces conceptually follows
the same procedure as the bulk case (Eq.~\ref{eqn:Eform}), and encounters the same
convergence issue due to periodic interactions of the charge, albeit now in
two dimensions instead of all three.
Fortunately, the correction scheme of Ref.~\citenum{PRLbulk}
generalizes for defects at surfaces as well.\cite{Komsa2013}
Intuitively, the only difference is that the self energy and potential
of the model charge are calculated using a dielectric slab $\epsilon_0(z)$
(or $\epsilon_\infty(z)$ for fixed atoms) to mimic the remaining material,
instead of a uniform dielectric covering all space for the bulk case.
See Ref.~\citenum{Komsa2013} for a detailed exposition;
we follow the same procedure except for two refinements.
First, we improve the electrostatic potential using Eq.~\ref{eq:Vnew} as before, which we discuss below.
Additionally, we develop a more robust and general method for evaluating
the isolated self-energy of the Gaussian charge model with an arbitrary
$\epsilon(z)$ background using a spectral expansion in
cylindrical Bessel functions (see SI),
instead of the image charge series method of Ref.~\citenum{Komsa2013}.

\begin{figure} 
\includegraphics[width=\columnwidth]{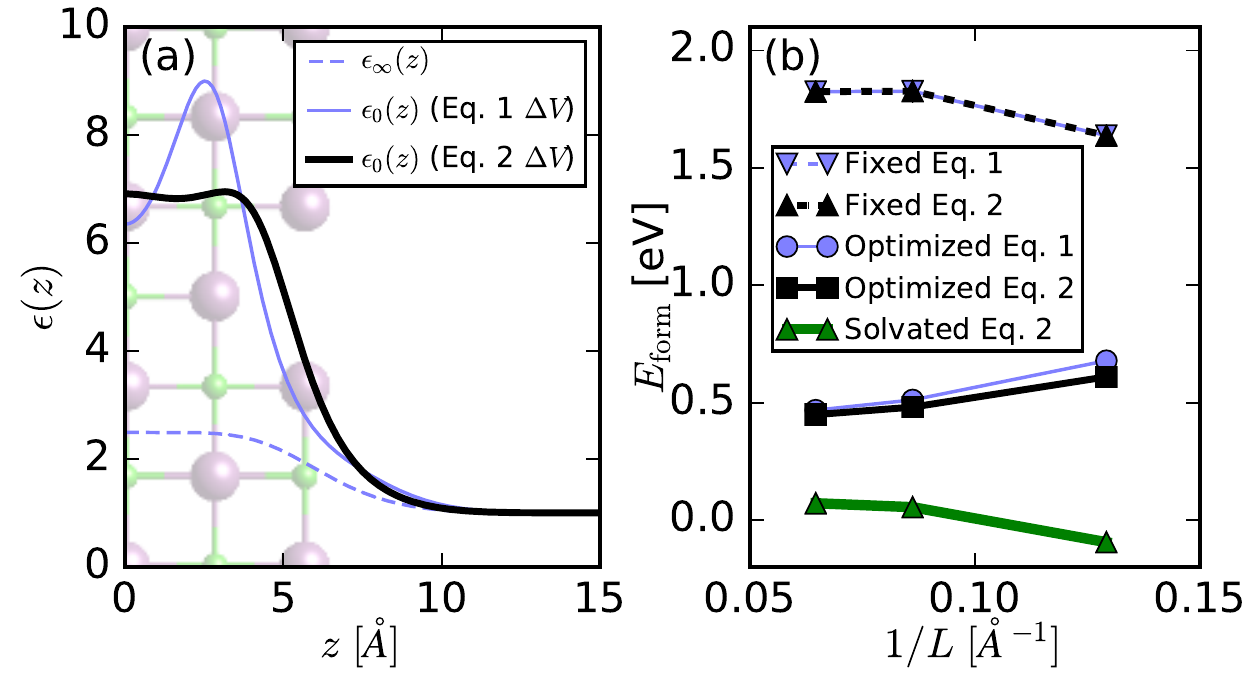}
\caption{(a) Dielectric profiles of 5-layer NaCl(100) slab calculated with fixed atom positions ($\epsilon_\infty(z)$)
and optimized atom positions ($\epsilon_0(z)$), using either Eq.~\ref{eq:Vorig} or 2 for the electrostatic potential.
The background shows the structure of this slab with a Cl vacancy (not included in dielectric calculation) to scale.
(b) Convergence of formation energy of a $q=+1$ Cl vacancy at the surface of this slab
with inverse supercell size, where $L = A^{1/2}$ for lateral supercell area $A$.
Eq.~\ref{eq:Vnew} improves convergence, but less dramatically so than in the bulk case.
Note that geometry optimization and solvation have large effects on vacancy formation energies.
\label{fig:SurfaceVacancy}}
\end{figure}

In the surface case, the electrostatic potential plays an additional role
in the determination of the dielectric slab model $\epsilon_0(z)$ (or $\epsilon_\infty(z)$).
Specifically, we apply a uniform normal electric field $E_0$ to the DFT calculation of the slab,
measure the change in the total electrostatic potential $\Delta V(z)$, and from that
calculate the dielectric function using $\epsilon^{-1}(z) = (1/E_0) (-\partial \Delta V(z)/\partial z)$.
This yields $\epsilon_0^{-1}$ or $\epsilon_\infty^{-1}$ depending on whether the atoms are optimized or fixed.
Fig.~\ref{fig:SurfaceVacancy}(a) shows the calculated dielectric profiles from potentials
using Eqs. 1 and 2 for an inversion symmetric 5-layer NaCl(100) slab.
By definition, the revised scheme does not change $\epsilon_\infty(z)$ since the atoms are fixed.
For $\epsilon_0(z)$, Eq.~\ref{eq:Vorig} produces oscillations with the lattice planes, despite the Gaussian smoothing,
while Eq.~\ref{eq:Vnew} produces a smooth transition from 1 outside the slab to the bulk value inside it.

Fig.~\ref{fig:SurfaceVacancy}(b) shows the corresponding formation for $q=+1$ Cl vacancies.
Once again, geometry optimization reduces the formation energy by a large amount ($\approx 1.3$~eV).
However, in this case, the effect of the smoother electrostatic potential using Eq.~\ref{eq:Vnew}
produces a less drastic improvement in the convergence with supercell size.
The primary reason is that the convergence with lateral supercell size is already reasonable
with Eq.~\ref{eq:Vorig}, because in this case $\Delta V$ between DFT and the model charge can
be aligned accurately in the vacuum region (outside the slab)
which is not affected by the large oscillations due to atom displacements.
Also note that unlike previous studies,\cite{Komsa2013} we do not need to
worry about convergence with the length of the normal direction because
we use truncated Coulomb potentials to exactly eliminate periodic interactions
in that direction for all calculations (neutral and charged).\cite{TruncatedEXX}

Finally, our methods for solvation,\cite{CANDLE} electrostatic potential evaluation
and charged defect correction for arbitrary $\epsilon(z)$ models (see SI) make it possible
now to straightforwardly predict formation energy of charged defects at solid-liquid interfaces.
Intuitively, this only requires using the solvation model in both the energy calculations in Eq.~\ref{eqn:Eform},
and including the solvent response contributions to the Gaussian charge model calculations.
Now, the net $\epsilon(z)$ includes the first-principles calculation for the solid slab
as above, plus an additional solvent contribution $(\epsilon_b - 1)s(z)$,
where $\epsilon_b$ is the bulk dielectric constant of the solvent in the solvation model
and $s(z)$ is the (planarly-averaged) cavity shape function $\in [0,1]$ that specifies
the distribution of solvent density assumed by the solvation model.\cite{CANDLE}
The solvent can additionally include a response $\kappa^2(z) = s(z)\frac{8\pi NZ^2}{k_BT}$
due to a concentration $N$ of ions of charge $Z$ in the electrolyte.
The generalized scheme presented in the SI straightforwardly handles such combinations
of $\epsilon(z)$ and $\kappa^2(z)$ for defect formation energy calculations.
Fig.\ref{fig:SurfaceVacancy}(b) shows that our method continues to yield excellent supercell size convergence
for the predicted formation energies of $q=+1$ Cl vacancies at an NaCl(001) surface in water (with 1M NaCl ions).
Note that the formation energies are strongly stabilized by about 0.5~eV relative to the vacuum surface.
 
\begin{figure}
\includegraphics[width=0.8\columnwidth]{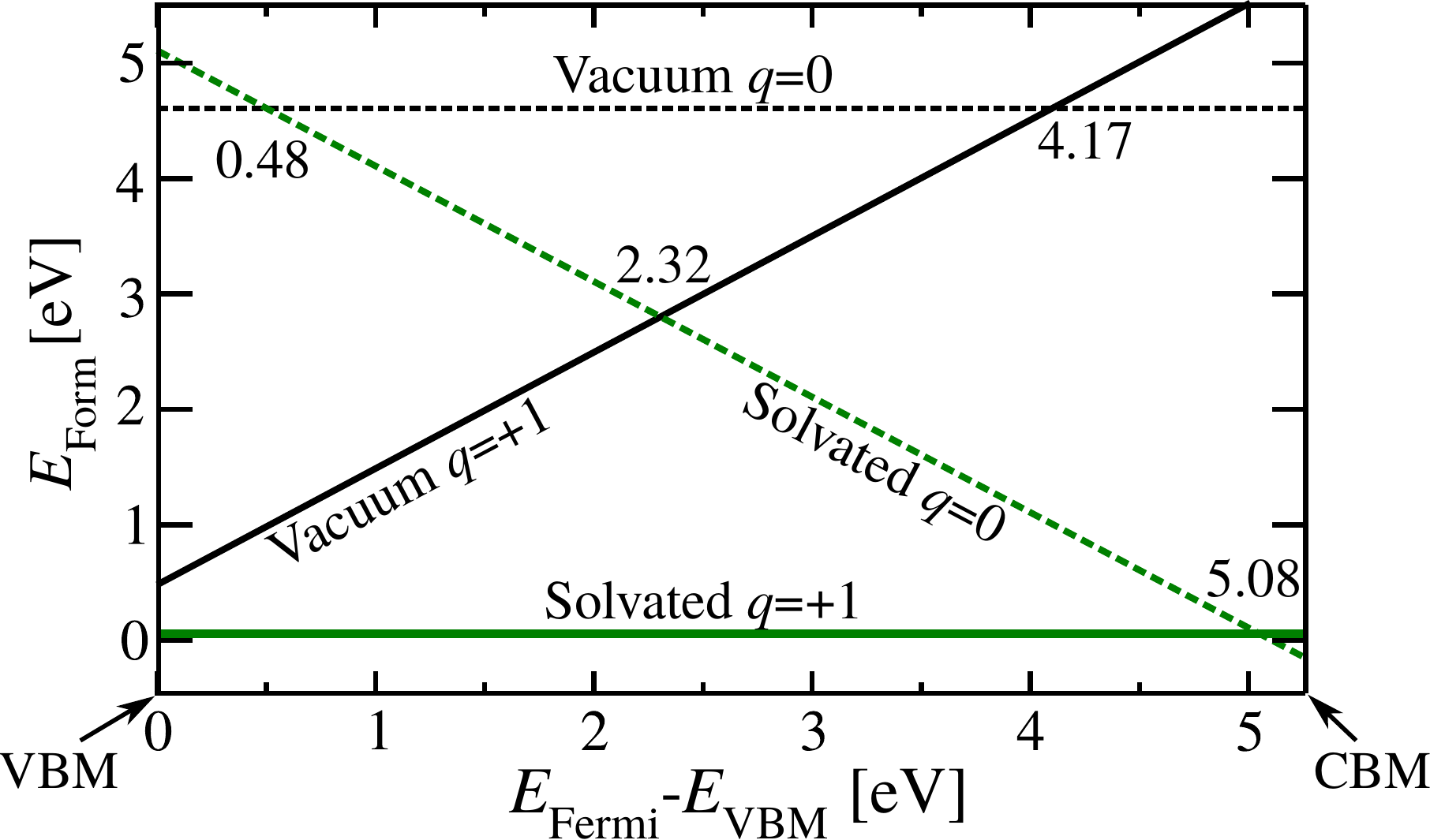}
\caption{Formation energies of neutral ($q=0$) and charged ($q=+1$) Cl vacancies
at an NaCl (001) surface in vacuum and solution, as a function of the Fermi level.
Charged vacancies are more stable for most Fermi level positions in both cases,
and vacancy formation is more favorable in solution overall.
\label{fig:interface-defect}}
\end{figure}

An important qualitative difference also arises due to the choice of
reference chemical potential of Cl, $\mu\sub{Cl}$, in Eq.~\ref{eqn:Eform}.
In vacuum, the natural choice is set by molecular/gaseous Cl\sub{2},
while in water, the natural choice is Cl$^{-1}$ ions in solution.
Fig.~\ref{fig:interface-defect} shows the variation of formation energies
of neutral and charged Cl vacancies at NaCl(001) surfaces in vacuum
and water with the Fermi level (or electron chemical potential).
Note that the charged vacancy in water and neutral vacancy have zero slope;
this is because without exchanging electrons, a Cl$^-$ ion can leave the
surface into solution, while a Cl atom can go to Cl\sub{2} gas in vacuum.
In both cases, the charged vacancy is more stable for most values of $E\sub{Fermi}$
in the band gap, and the overall formation energies of vacancies
are much smaller in solution than in vacuum.

\section*{Conclusions}

In summary, we show that a simple revised scheme of calculating
DFT electrostatic potentials by subtracting pseudized atom potentials
suppresses oscillations in these potentials by over an order of magnitude.
This makes it possible to efficiently and accurately predict band alignments at interfaces,
and energetics of charged defects in solids, at surfaces and even at solid-liquid interfaces,
with much smaller slab and supercell sizes.
The substantial stabilization of charged defects at solid-liquid interfaces underscores
the importance of evaluating charged defective surfaces in solution.
While this was previously formidable due to the difficulty of dealing with
electrostatic potential alignment in expensive molecular dynamics simulations, the combination
of our revised scheme with continuum solvation theories make such calculations now practical.

\section*{Supplemental Material}

Derivation of a general method to calculate the self energy of a Gaussian charge model
at an arbitrary planar interface with dielectric and/or Debye response.
This is useful for charged defect formation energy calculations at such interfaces.

\makeatletter{} 
\end{document}